\documentclass[twocolumn,aps,floatfix,showpacs,superscriptaddress]{revtex4}
\usepackage{amsmath,amssymb,eucal,graphicx}

\begin{document}
\title{Flows on Graphs with Random Capacities}

\author{T. Antal}\affiliation{Program for Evolutionary Dynamics, Harvard
  University, Cambridge, Massachusetts 02138, USA}
\author{P.~L.~Krapivsky} \affiliation{Department of Physics, Boston
  University, Boston, Massachusetts 02215, USA}

\begin{abstract}
We investigate flows on graphs whose links have random capacities. 
For binary trees we derive the probability distribution for the maximal flow from 
the root to a leaf, and show that for infinite trees it vanishes beyond a 
certain threshold that depends on the distribution of capacities. We then examine 
the maximal total flux from the root to the leaves. Our methods generalize to simple graphs with loops, {\it e.g.}, to hierarchical lattices and to complete graphs.
\end{abstract}
\pacs{02.50.Cw, 05.60.-k, 89.40.-a, 89.75.Hc}


\maketitle

\section{introduction}

Flows in networks are abundant: the flow of water in rivers and pipelines, the flow of current in electrical wires, the flow of passengers, the flow of cars through the network of roads \cite{traffic} are just a few examples. These flows can be characterized by the conservation of current -- apart from sources and sinks in the network, the current is locally conserved.

The maximal flows in capacitated networks are especially interesting \cite{network}. A capacitated network is a graph with a non-negative number called capacity $c=c(e)$ assigned to each edge $e$. Capacity measures the maximal flow that can pass through the edge. For each vertex, the current flowing in and out of it should be the same (Kirchoff's conservation law). This rule is only modified for the source vertices where the current enters the system, and the sink vertices through which the current leaves the network. Practically it means, that for two edges $e_1,e_2$ in series the combined capacity of the path $(e_1,e_2)$ is $\min(c_1,c_2)$; for two edges in parallel the combined capacity is $c_1+c_2$. 

\begin{figure}
\centering
\includegraphics[scale=0.13]{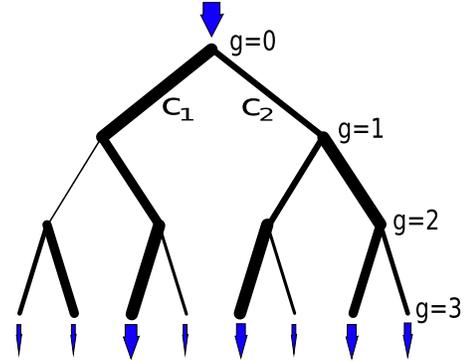}
\caption{Illustration of a flow on a binary tree of three generations. The width of the links represents their capacity, for instance $c_1>c_2$. The current flows from top to bottom. The arrows represents the amount of current flowing in and out of the system.}
\label{illu}
\end{figure}

In this paper we mainly focus on rooted trees in which all leaf nodes are at the same distance from the root; such trees are often called perfect trees.  
We assume that the current enters through 
the root of the tree and it is discharged at the leaves (see Fig.~\ref{illu}).
Generally for a path $(e_1,\ldots,e_n)$, {\it i.e.}, a set of edges in series, the capacity of a path is the minimal capacity along the path: $\min(c_1,\ldots,c_n)$. This defines the flow between any two vertices in a tree.

There is a huge engineering and mathematical literature on flows in networks. One important goal is to find the maximal flow \cite{shannon,ford,network,rivest,lee}, 
although other issues have been also investigated, see e.g. 
\cite{fisher,redner,cieplak,mass,congestion,satya,sam,mossel,durand,examp}. Here we concentrate on large networks where the capacities are random variables chosen from the same distribution. We shall mostly investigate trees since the absence of circuits greatly simplifies the analysis. In Sec.~\ref{bintree}, we investigate flows on deterministic rooted trees, namely on binary trees and then on $b-$ary trees. In Sec.~\ref{hierlat}, we solve the maximal flow problem on hierarchical lattices which generalize rooted trees to include loops. We then analyze random recursive trees in Sec.~\ref{RRT}. A brief discussion is given in Sec.~\ref{disco}.

\section{Deterministic Rooted Trees}
\label{bintree}

A rooted binary tree, also known as a Cayley tree of coordination number 3, has a root vertex in generation $g=0$ which is joined to 2 vertices in generation $g=1$ each of which is joined to 2 vertices in generation $g=2$, so that there are a total 4 vertices in generation 2; generally, the binary tree has $2^g$ vertices in generation $g$ 
(Fig.~\ref{illu}). Suppose that current can flow only in one direction, namely from 
generation $g$ to $g+1$. A capacity $c=c(e)\ge0$ is assigned randomly to 
each edge $e$. We shall assume that capacities are independently chosen from a distribution with density $f(s)=-F'(s)$ where 
\begin{equation}
\label{def:f}
F(s)={\rm Prob}\{c>s\} ~.
\end{equation}

For the binary tree with $g$ generations, there are $2^g$ leaves (vertices in the last generation $g$).  The capacity $\min(c_1,\ldots,c_g)$ of each path from the root to the leaf is a random variable whose distribution is given by
\begin{equation*}
\begin{split}
{\rm Prob}\{\min(c_1,\ldots,c_g)> s\}&=\\
\prod_{k=1}^g {\rm Prob}(c_k > s) &= [F(s)]^g ~.
\end{split}
\end{equation*}
What is the probability distribution of the maximal flow among these $2^g$ flows? 
What is the distribution of the total flux from the root to the leaves? 
These are the questions discussed below. 

\subsection{Maximal flow}
\label{bin-max}

Let $M_g = \max(s_1,\dots,s_{2^g})$ be the maximal flow out of the $2^g$ flows to all  leaves. Let
\begin{equation}
\label{def:M}
P_g(s)={\rm Prob}\{M_g\leq s\} ~
\end{equation}
be the cumulative distribution.
Each path goes through one of the two edges issuing from the root
(we denote their capacities by $c_1$ and $c_2$ as illustrated on Fig.~\ref{illu}). Therefore 
the probability $P_g(s)$ is given by
\begin{equation*}
{\rm Prob}\{\min(c_1,M_{g-1}^{(1)})\le s\}
\times{\rm Prob}\{\min(c_2,M_{g-1}^{(2)})\le s\}
\end{equation*}
where $M_{g-1}^{(1)}, M_{g-1}^{(2)}$ are the maximal flows in the corresponding
daughter trees (whose roots are vertices from generation one).  The daughter
trees are statistically independent and hence $M_{g-1}^{(1)}, M_{g-1}^{(2)}$
have the same distribution. Since
\begin{equation*}
{\rm Prob}\{\min(c,M_{g-1}) > s\}=F(s)\,[1-P_{g-1}(s)]
\end{equation*}
we arrive at a recurrence
\begin{equation}
\label{rec}
P_g(s)=\{1-F(s) + F(s)P_{g-1}(s)\}^2 ~.
\end{equation}
This recurrence could have been derived in many other ways. 
To have a smaller than $s$ maximal flow in the tree, one needs, independently in both sub-trees, either the upmost capacity to be smaller than $s$ \{term $1-F(s)$\},
or if it is larger than $s$,  one needs the maximal flow in the daughter tree to be smaller than $s$ \{term $F(s)P_{g-1}(s)$\}.

\begin{figure}
\centering
\includegraphics[scale=0.7]{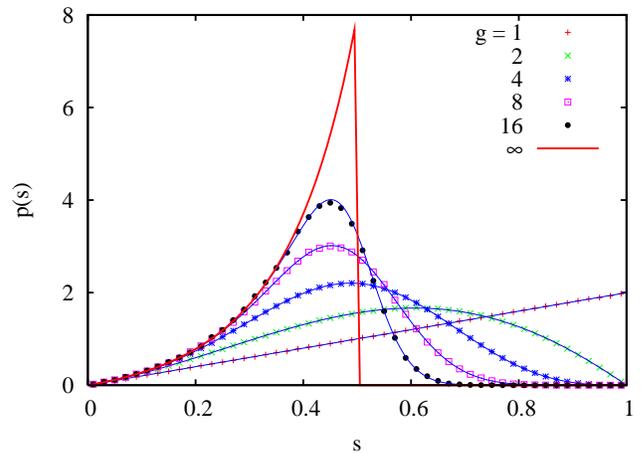}
\caption{Density $p(s)$ of the maximal flow though a binary tree of $g$ generations. The capacities were drawn independently from a flat distribution. Simulation results and analytic solution is depicted for finite $g$ values. The $g\to\infty$ asymptotic limit is also shown.}
\label{locflat}
\end{figure}

Iterating (\ref{rec}) one gets
\begin{eqnarray*}
P_1&=& (1-F)^2\\
P_2&=&\left[1-F+F(1-F)^2\right]^2\\
P_3&=&\left\{1-F+F\left[1-F+F(1-F)^2\right]^2 \right\}^2
\end{eqnarray*}
etc. The expressions $P_g(s)$ become unwieldy but quickly approach a limiting
distribution $P_\infty(s)=\lim_{g\to\infty}P_g(s)$ which has a remarkably
simple form
\begin{equation}
P(s) = \left\{    
\begin{array}{cl} \displaystyle
\left[ \frac{1-F(s)}{F(s)}\right]^2 & \mbox{~~for~~} 1/2< F(s) \le 1\\ \\
\displaystyle
1 &  \mbox{~~for~~} 0\leq F(s)\leq 1/2 ~.
\end{array}
\right. 
\end{equation}
(Hereinafter we will omit the $g=\infty$ subscript.) The corresponding density function is
\begin{equation}
p(s) = \left\{    
\begin{array}{cl} \displaystyle
\frac{ 2f(s)[1-F(s)]}{[F(s)]^3} & \mbox{~~for~~} F(s)> 1/2\\ \\
\displaystyle
0 &  \mbox{~~for~~} F(s)<1/2 ~.
\end{array}
\right. 
\end{equation}

The most intriguing feature of this density function is the sharp cutoff at $s^*$, which is given by $F(s^*)=1/2$. Note that the position of the cutoff $s^*$ does not depend on the details of the capacity distribution, only on the location it takes the value $1/2$. In other words, the specific values of the large capacities ($s$ with $F(s)<1/2$) do not matter  for large trees. For instance, if the capacity distribution is bimodal, $f(s)=(1-a)\delta(s)+a\delta(b-s)$, then when $a<1/2$ the maximal current is zero independent of $b$ while for $a>1/2$ the maximal current is $b$. This is essentially a branching process, where the process survives if the branching parameter is larger than $1/2$.

We now illustrate the behavior of the limiting distribution in two representative cases. As an example of a bounded capacity distribution, we choose  the flat distribution with density $f(s)=1$ if $0\leq s\leq 1$ and $f(s)=0$ when $s>1$. Then 
\begin{equation}
\label{p}
p(s) = \left\{    
\begin{array}{cl} \displaystyle
\frac{2s}{(1-s)^3} & \mbox{~~for~~} s<1/2\\ \\
\displaystyle
0 &  \mbox{~~for~~} s>1/2 ~.
\end{array}
\right. 
\end{equation}
The sharp cutoff in the limiting distribution, and also the relatively slow convergence of the finite generation curves can be observed on Fig.~\ref{locflat}. {}From \eqref{p} one can compute any moment of the maximal flow. For instance, the average maximal flux is $\langle s\rangle=\int ds\,sp(s)=2\ln 2 -1$.

As an example of an unbounded capacity distribution we take the exponential distribution, $f(s)=e^{-s}$. Then 
\begin{equation}
\label{p-exp}
p(s) = \left\{    
\begin{array}{cl} \displaystyle
2e^s(e^s-1) & \mbox{~~for~~} s<\ln 2\\ \\
\displaystyle
0 &  \mbox{~~for~~} s> \ln 2 
\end{array}
\right. 
\end{equation}
and the average maximal flux is $\langle s\rangle=1/2$. 

\subsection{Total flux}
\label{bin-tot}

Consider the total flux (more precisely the maximal possible total flux) from the root to the leaves. For the binary tree with one generation, the total flux from the root
to the two leaves is $\Phi_1=c_1+c_2$. Similarly for the binary tree with two generations  
\begin{equation*}
\Phi_2=\min[c_1,c_{11}+c_{12}]+\min[c_2,c_{21}+c_{22}] ~.
\end{equation*}
Generally the total flux from the root to the bottom of the binary tree is found from 
\begin{equation}
\label{sumrule}
\Phi_{g}=\min[c_1,\Phi_{g-1}^{(1)}]+\min[c_2,\Phi_{g-1}^{(2)}] ~,
\end{equation}
where $\Phi_{g-1}^{(1)}, \Phi_{g-1}^{(2)}$ are the total fluxes in the corresponding
daughter trees (whose roots are vertices from generation one).  

Let $R_g(q)={\rm Prob}(\Phi_g>q)$ and $\rho_g(q)=-R_g'(q)$ be the corresponding density function of the total flux. As the daughter trees are statistically independent, 
equation \eqref{sumrule} implies that
\begin{equation}
\label{conv-g}
\rho_g(q) = \int_0^q dx\,h_g(x)h_g(q-x)
\end{equation}
where $h_g$ is the probability density of the total flux of a half tree 
\begin{equation*}
h_g(q)=-H_g'(q)\,,\quad H_g(q) = {\rm Prob}\{\min[c,\Phi_{g-1}] >q\}
\end{equation*}
Since the minimum of two numbers are larger then $q$ if and only if both numbers are larger than $q$, we have 
\begin{equation}
\label{r-gen-g}
H_g(q)= F(q)\,R_{g-1}(q) ~.
\end{equation}
Setting $R_0(q)\equiv 0$, Eqs.~\eqref{conv-g}--\eqref{r-gen-g} provide a recursive formula for $\rho_g(q)$ for any finite $g$. In the $g\to\infty$ limit we arrive at
\begin{subequations}
\begin{align}
\label{conv}
    \rho(q) &= \int_0^q dx\,h(x)h(q-x)\\
\label{r-gen}
    H(q) &=F(q)\,R(q) ~.
\end{align} 
\end{subequations}
Equation \eqref{conv} shows that $\rho(q)$ is a convolution. This suggests
to employ the Laplace transform, which we denote by a hat $\hat {\mathcal F}(\lambda)=\int_0^\infty dq\,e^{-\lambda q}\, {\mathcal F}(q)$. We can recast \eqref{conv} into
\begin{equation}
 \label{Lapgen}
\hat\rho(\lambda)=[\hat h(\lambda)]^2 = [1-\lambda \hat{H}(\lambda)]^2
\end{equation}
and hence, with \eqref{r-gen}, we have a closed equation for the Laplace transform of the flux density.

First, let us discuss the properties of $\rho(q)$ for a general capacity density $f(q)$. Obviously, $\rho(q)$ has a finite support if and only if $f(q)$ has a finite support. From \eqref{conv} and \eqref{r-gen} we find linear small $q$ behavior $\rho(q) \to qf^2(0)$ in the generic case $f(0)>0$. In Appendix \ref{smallq} we obtain the complete small $q$ series of $\rho(q)$. 

We now again consider two representative examples.
For the exponential distribution of capacities, Eq.~\eqref{Lapgen} becomes
\begin{equation}
\label{Lap}
\hat\rho(\lambda)=\left[\frac{1+\lambda \hat\rho(\lambda+1)}{1+\lambda}\right]^2 ~.
\end{equation}
By iteration, one arrives at a formal solution
\begin{equation*}
\hat\rho\!=\!\Big(\frac{1}{1+\lambda}+\frac{\lambda}{1+\lambda}
\Big(\frac{1}{2+\lambda}+\frac{1+\lambda}{2+\lambda}
\Big(\frac{1}{3+\lambda}+\ldots\Big)^2\Big)^2\Big)^2
\end{equation*}

It seems hard to recast the Laplace transform into a more compact form, and to invert the Laplace transform. One can extract, however, the asymptotics of the flux density already from Eq.~(\ref{Lap}). The small $q$ behavior of $\rho(q)$ is encoded in the large $\lambda$ behavior of the Laplace transform. Using Eq.~(\ref{Lap}) one derives the large $\lambda$ expansion of the Laplace transform
\begin{equation*}
\hat\rho(\lambda)=\lambda^{-2}-4\lambda^{-4}+ 2\lambda^{-5} + 30\lambda^{-6} + \ldots
\end{equation*}
from which
\begin{equation}
\label{small-x}
\rho=q-\frac{2}{3}\,q^3+\frac{1}{12}q^4+\frac{1}{4}q^5 + \ldots\quad {\rm as}\quad q\to 0.
\end{equation}
This series can be also obtained from the general formula given in Appendix \ref{smallq}.

If the large $q$ behavior of $\rho(q)$ were exponential, it would be
encoded in the poles of its Laplace transform. We now show that the Laplace 
transform $\hat\rho(\lambda)$ is an entire function, {\it i.e.}, it has no poles or other singularities in the complex $\lambda$ plane. An apparent pole at $\lambda=-1$ is not a pole --- using the expansion
\begin{equation}
\label{small-lambda}
\hat\rho(\lambda)=1-\lambda\langle q\rangle+\mathcal{O}(\lambda^2)
\end{equation}
one finds that $\hat\rho(-1)=(1+\langle q\rangle)^2$. Similarly 
one computes $\hat\rho(-2)=(1+4\langle q\rangle+2\langle q^2\rangle)^2$
and $\hat\rho(N)$ for other negative integers $N$ and finds that all the
apparent poles are regular points.  The lack of poles means that $\rho(x)$ decays faster than exponentially.

\begin{figure}
\centering
\includegraphics[scale=0.7]{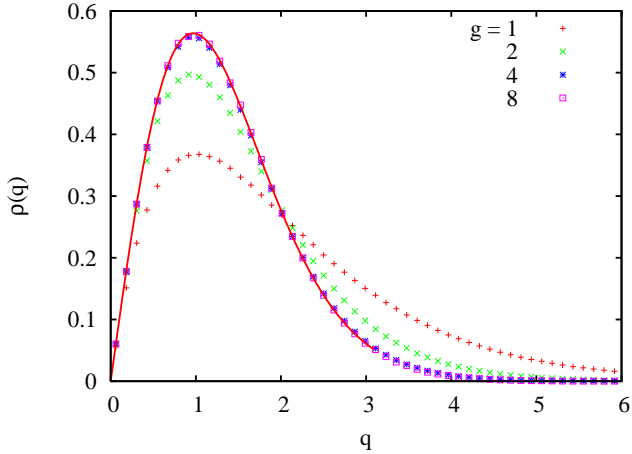}
\caption{Simulation results for the density $\rho(q)$ of the total flux though a binary tree of $g$ generations. The capacities were drawn independently from an exponential distribution. The solid line represents the small $q$ asymptotic solution given in 
Appendix \ref{smallq}.}
\label{totexpo}
\end{figure}

To establish the decay law we use equation (\ref{Lap}) to find
\begin{equation}
\label{v}
\hat\rho(\lambda)\propto \exp\left\{v\,2^{-\lambda}\right\}
\end{equation}
for large negative $\lambda$ from which 
\begin{equation}
\label{rq-large}
\rho(q)\propto \exp\left\{-\frac{q\ln q}{\ln 2} -q\,\frac{1+\ln v}{\ln 2}\right\}
\end{equation}
as $q\to\infty$. 
Thus $\rho(x)$ decays faster than exponentially. The leading factorial asymptotic 
$\rho\propto e^{-q\log_2 q}$ is independent on $v$, while the exponential correction 
depends on $v$ whose determination requires the analysis of the full equation  
(\ref{Lap}). The asymptotic predictions \eqref{small-x} and \eqref{rq-large} are in good agreement with simulation results (Fig.~\ref{totexpo}). 

\begin{figure}
\centering
\includegraphics[scale=0.7]{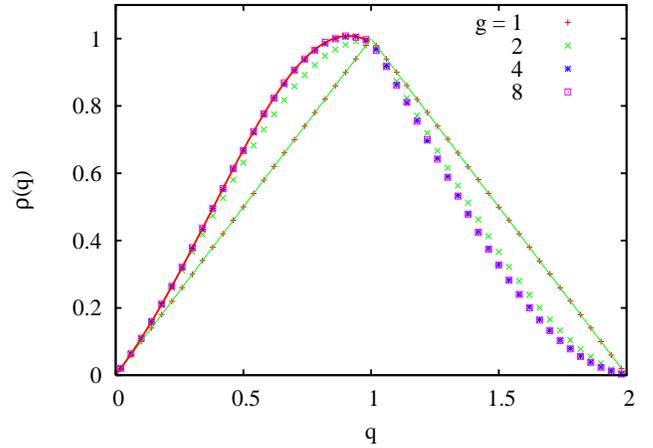}
\caption{Simulation results for the density $\rho(q)$ of the total flux though a binary tree of $g$ generations. The capacities were drawn independently from a flat distribution. The distribution for the first generation is composed by two straight lines. The thick line represents the small $q$ asymptotic solution, valid for $q<1$ and given in 
Appendix \ref{smallq}.}
\label{totflat}
\end{figure}

For the flat capacity distribution, Eq.~\eqref{r-gen} gives 
\begin{equation}
\label{r}
h(q)=(1-q)\,\rho(q)+\int_q^2 dy\,\rho(y)\quad{\rm for}\quad q<1 
\end{equation}
and $h(q)=0$ for $q>1$.  
Equation \eqref{conv} shows that $\rho(q)$ vanishes outside the interval 
$0<q<2$, while inside the interval we utilize the fact that $h(q)=0$ when $q>1$ 
and re-write (\ref{conv})  as
\begin{equation}
\label{rho}
\rho(q) = \left\{    
\begin{array}{cl} \displaystyle
\int_0^q dx\,h(x)h(q-x) & \mbox{~~for~~} 0\leq q<1\\ \\
\displaystyle
\int_{q-1}^1 dx\,h(x)h(q-x)&  \mbox{~~for~~} 1<q\leq 2 ~.
\end{array}
\right. 
\end{equation}
The small $q$ expansion of the flux density is found from Eqs.~\eqref{r}--\eqref{rho} to yield
\begin{equation}
\label{small-q}
\rho(q)=q+q^2-\frac{1}{6}\,q^3-\frac{5}{6} q^4 + \ldots
\end{equation}
in agreement with the general formula given in Appendix \ref{smallq}. 
Similarly by expanding Eqs.~\eqref{r}--\eqref{rho} near the upper cutoff we find that the 
flux density vanishes linearly according to
\begin{equation}
\label{large-q}
\rho(q)\to (2-q)\mu^2\,,\quad \mu\equiv \int_1^2 dx\,\rho(x) ~.
\end{equation}
The normalization requirement $\int_0^2 dx\,\rho(x)=1$ shows that $\mu<1$, but to compute $\mu$ seems impossible without solving the entire problem. Numerically 
$\mu\approx 0.4$ and the asymptotic predictions \eqref{small-q} and \eqref{large-q} are in good agreement with simulation results (Fig.~\ref{totflat}). Interestingly, 
the total flux density converges remarkably fast in $g$, as opposed to the slow convergence of the maximal flow density shown on Fig.~\ref{locflat}. 

\subsection{$b$-ary trees}
\label{b}

The above framework can be extended to other kinds of trees. For instance, for the 
$b$-ary tree (also known as a rooted Cayley tree with coordination number $b+1$) each node (apart from the root) has one incoming link and $b$ outgoing links. The analog
of Eq.~\eqref{rec} reads 
\begin{equation}
\label{rec-r}
P_g(s)=\{1-F(s)+F(s)P_{g-1}(s)\}^b ~.
\end{equation}
Therefore the stationary distribution $P\equiv P_\infty$ 
satisfies $P=[1-F+FP]^b$. The  trivial root of this equation gives the limiting distribution for large $s$, 
\begin{equation*}
P(s)=1 \quad{\rm when} \quad F(s)<1-1/b
\end{equation*}
while for $F(s)\geq 1-1/b$, the limiting distribution is given by a non-trivial
root. For instance, for ternary trees
\begin{equation*}
P=\frac{2F^2-3F+\sqrt{4F-3F^2}}{2F^2} 
\end{equation*}
for $F(s)> 2/3$. In particular, for the ternary tree with a flat
distribution of capacities
\begin{equation*}
P = \left\{    
\begin{array}{cl} \displaystyle
\frac{1}{2}\left[\sqrt{\frac{1+3s}{(1-s)^3}}+\frac{3s-1}{1-s}\right] 
& \mbox{~~for~~} s<1/3\\ \\
\displaystyle
1 &  \mbox{~~for~~} s\ge 1/3 ~.
\end{array}
\right. 
\end{equation*}

Further, the distribution of the total flux $\Phi\equiv \Phi_\infty$  in the infinite rooted $b$-ary tree is invariant under the transformation
\begin{equation}
\label{invariant}
\Phi=\min[c_1,\Phi]+\ldots+\min[c_b,\Phi] ~.
\end{equation}
Therefore the flux density is a (generalized) convolution
\begin{equation}
\label{rho-b}
\rho(q) = \int_0^q\ldots\int_0^q \prod_{j=1}^b dx_j\,h(x_j)\,\,\delta(q-x_1-\ldots-x_b)
\end{equation}
and the Laplace transform is $\hat\rho(\lambda)=[\hat h(\lambda)]^b$. As an example,
let us consider the exponential distribution. Then the Laplace transform of the flux density becomes 
\begin{equation}
\label{Lap-b}
\hat\rho(\lambda)=\left[\frac{1+\lambda \hat\rho(\lambda+1)}{1+\lambda}\right]^b ~.
\end{equation}
The small $q$ expansion is 
\begin{equation}
\label{small-b}
\rho=\frac{q^{b-1}}{(b-1)!}+\mathcal{O}(q^b)
\end{equation}
while when $q\to\infty$ the flux density decays according to 
\begin{equation}
\label{large-b}
\rho(q)\propto e^{-q\log_b q} 
\end{equation}

\section{Hierarchical lattices}
\label{hierlat}

Hierarchical lattices represent a simple generalization of rooted trees. These lattices have loops but are still tractable (see e.g. \cite{reno,derrida,slow}). They mimic finite dimensional lattices. 

\begin{figure}
\centering
\includegraphics[scale=0.13]{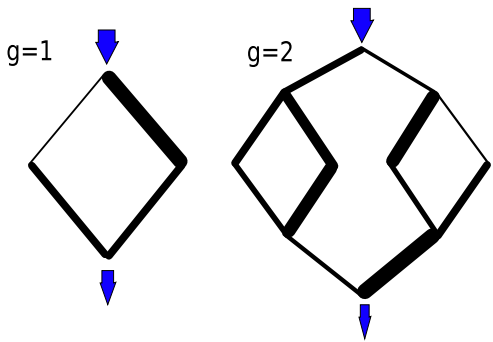}
\caption{Illustration of a flow on a binary hierarchical lattices of one and two generations. The width of the links represents their capacity, and the arrows stand for the amount of current flowing in and out of the system.}
\label{illu_hier}
\end{figure}

A $b$-ary hierarchical lattice of $g$ generations is composed by two $b$-ary trees of $g$ generations merged at the leaves. The current enters the system at one root, and leaves it at the other. In Fig.\ref{illu_hier} we show the binary hierarchical lattices of one and two generations. 

Denoting again the flux distribution of a half of the lattice by $H_g$ we get
\begin{equation}
\label{rl-gen-g}
H_g(q)= [F(q)]^{2}R_{g-1}(q) ~.
\end{equation}
In contrast to Eq.~\eqref{r-gen-g} we have $F^{2}$ since the capacities of {\em two} edges attached to the daughter lattice with $g-1$ generations should be smaller 
than $q$. 

The total flux is given by a $b$-fold convolution analogous to \eqref{rho-b}, which again simplifies after performing the Laplace transform
\begin{equation}
 \label{Lapgen_hier}
\hat\rho_g(\lambda)=[\hat h_g(\lambda)]^b = [1-\lambda \hat{H_g}(\lambda)]^b ~.
\end{equation}
A finite $g$ distribution can be obtained by starting with $R_0(q)\equiv 0$ and
iterating the above equations. 

The limiting flux distribution is determined by solving
\begin{subequations}
\begin{align}
\label{HFR}
   H(q)&= [F(q)]^{2}R(q)\\
\label  {rh} 
   \hat\rho(\lambda)&=[1-\lambda \hat{H}(\lambda)]^b 
\end{align} 
\end{subequations}
In the generic case of $f(0)>0$ we get
\begin{equation}
\label{rho-b-small}
\rho(q)=\frac{[2f(0)]^b}{(b-1)!}\,q^{b-1}+\mathcal{O}(q^b)
\end{equation}
in the small $q$ limit. 

\begin{figure}
\centering
\includegraphics[scale=0.7]{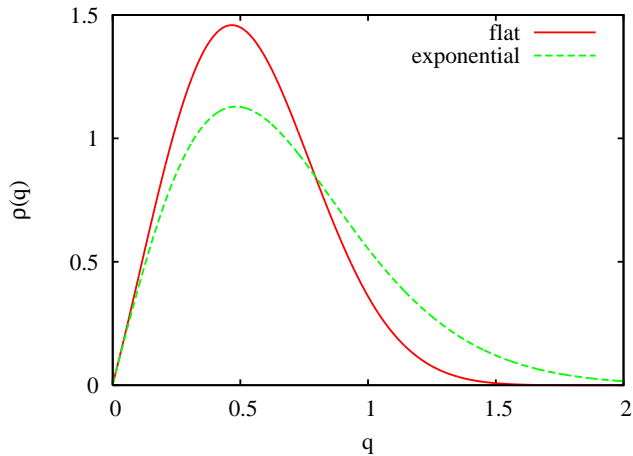}
\caption{Total flux density $\rho(q)$ for binary hierarchical lattice in the infinite size limit ($g\to\infty$). The capacities were drawn from a flat and from an exponential distribution. The curves were obtained by iterating \eqref{rl-gen-g}, and recovered by the small $q$ asymptotic series given in Appendix \ref{smallq}. }
\label{hier}
\end{figure}

The large $q$ behavior is less universal. Let us consider again two representative cases exemplifying unbounded and bounded capacity distributions. For the exponential capacity distribution, Eqs.~\eqref{HFR}--\eqref{rh} give
\begin{equation}
\label{Lap-b-exp}
\hat\rho(\lambda)=\left[\frac{2+\lambda \hat\rho(\lambda+2)}{2+\lambda}\right]^b ~.
\end{equation}
Using this expression we extract the asymptotic behaviors. We find that the small 
$q$ expansion agrees with the general asymptotic \eqref{rho-b-small} while 
when $q\to\infty$ the flux density decays according to 
\begin{equation*}
\rho(q)\propto e^{-2q\log_b q} 
\end{equation*}
These predictions agree with simulation results (Fig.~\ref{hier}).

For the flat capacity distribution we arrive [by differentiating Eq.~\eqref{HFR}] at an integral equation
\begin{equation}
\label{r-flat}
h(q)=(1-q)^2\rho(q)+2(1-q)\int_q^2 dy\,\rho(y)
\end{equation}
for $q<1$. 
Another relation between functions $\rho(q)$ and $h(q)$ is 
(in the case of the binary hierarchical lattice) given by Eq.~\eqref{rho}. 

To extract the behavior of $\rho(q)$ in the 
$q\to 2$ limit we write $q=2-\epsilon, x=1-\epsilon y$ and re-write Eq.~\eqref{rho} as
\begin{equation}
\label{r2}
\rho(q) = \epsilon\int_0^1 dy\,h[1-\epsilon y]\,h[1-\epsilon (1-y)]
\end{equation}
{}From Eq.~\eqref{r-flat} we get
\begin{equation}
\label{h1}
h[1-\epsilon z]=2\epsilon \mu z+\mathcal{O}(\epsilon^2)
\end{equation}
where again $\mu=\int_1^2 dx\,\rho(x)$. Plugging \eqref{h1} into Eq.~\eqref{r2} and computing the integral we find that $\rho(q)$ vanishes cubically near the upper cutoff,
\begin{equation}
\rho(q)=\frac{2}{3}\,\mu^2(2-q)^3+\mathcal{O}[(2-q)^4]
\end{equation}
Generally for the $b$-ary hierarchical lattice $\rho\propto (2-q)^{b+1}$ near the upper cutoff.

\section{Random recursive trees}
\label{RRT}

In the previous sections we have studied flows on regular graphs where all node degrees (apart from the root node) were the same. It is harder to deal with arbitrary random graphs but at least one class of random trees, the so-called random recursive trees, are tractable. First we overview the general properties of these random trees, and then we shall discuss flows on these trees. 

A random recursive tree is generated as follows: One starts with a root and adds nodes one by one so that each newly-introduced node is linked to a randomly selected existing node.  Random recursive trees which have been studied in great detail 
(see  e.g. the survey \cite{mh} and more recent articles \cite{drmota,leader,sj,fill}).

Let $N$ be the total number of nodes and $D_j(N)$ be the number of nodes on distance $j$ from the root.  By definition, $D_0(N)\equiv 1$, since the root is on distance 0 from
itself. Quantities $D_j(N)$ with $1\leq j<N$ are random, and e.g. the number of nodes on distance one from the root evolves according to 
\begin{equation}
\label{D1}
D_1(N+1)=
\begin{cases}
D_1(N)+1   & {\rm prob}\quad N^{-1}\cr
D_1(N)       & {\rm prob}\quad 1-N^{-1} ~.
\end{cases} 
\end{equation}
Thus the average 
$\mathcal{D}_1(N)\equiv \langle D_1(N)\rangle$ satisfies
\begin{equation*}
\mathcal{D}_1(N+1)=\mathcal{D}_1(N)+\frac{1}{N}
\end{equation*}
whose solution is $\mathcal{D}_1(N)=\sum_{1\leq j\leq N-1}j^{-1}$. Similarly
using (\ref{D1}) one establishes a simple recurrence for the variance  $\mathcal{V}_1(N)=\langle D_1^2(N)\rangle-\langle D_1(N)\rangle^2$, viz.
\begin{equation*}
\mathcal{V}_1(N+1)=\mathcal{V}_1(N)+\frac{1}{N}-\frac{1}{N^2}
\end{equation*}
which is also solvable. Asymptotically
\begin{subequations}
\begin{align}
&\mathcal{D}_1(N)=\ln N+\gamma+\mathcal{O}\left(\frac{1}{N}\right)
    \label{d1}\\
&\mathcal{V}_1(N)=\ln N+\gamma-\frac{\pi^2}{6}+\mathcal{O}\left(\frac{1}{N}\right) 
    \label{v1}
\end{align} 
\end{subequations}
where $\gamma=0.5772\ldots$ is Euler's constant.

Thus fluctuations are asymptotically negligible \cite{note} and we can focus on the averages. One easily establishes the exact recurrence for the averages
\begin{equation}
\label{Dj}
\mathcal{D}_j(N+1)=\mathcal{D}_j(N)+\frac{1}{N}\,\mathcal{D}_{j-1}(N) ~.
\end{equation}
In the $N\to\infty$ limit, this recurrence reduces to a differential equation
$\frac{d\mathcal{D}_j}{dN}=N^{-1}\,\mathcal{D}_{j-1}$ whose solution reads
\begin{equation}
\label{Dj-sol}
\mathcal{D}_j(N)=\frac{(\ln N)^j}{j!} ~.
\end{equation}

Thus the root is linked to approximately $\ln N$ nodes each of which in turn is linked on average to $\frac{1}{2}\,\ln N$ nodes. What is important is that each node in the first generation (on distance one from the root)  is linked to many nodes in the second generation; only a few nodes in the first generation have a finite (not diverging with $N$) number of links to the nodes of the second generation. This observation allows us to compute the maximal flow in the leading order. Indeed, asymptotically it is the maximal capacity among all $D_1(N)$ capacities from the root to the first generation nodes. 

The above argument shows that for the flat distribution of capacities the density distribution for the maximal flow approaches a stationary limit $p_\infty(s)=\delta(s-1)$. Essentially the same behavior occurs for any bounded capacity distribution: 
$p_\infty(s)=\delta(s-s_{\rm max})$. For unbounded capacity distributions, the density distribution for the maximal flow does not approach a stationary limit. For instance, for the exponential distribution we approximately have
\begin{equation}
\label{pN}
p_N(s)=\ln N e^{-s}(1-e^{-s})^{\ln N}
\end{equation}
implying that on average the maximal flow is
\begin{equation}
\label{sN}
\langle s\rangle_N=\ln N ~.
\end{equation}

The computation of the total flux is even simpler. For an arbitrary capacity distribution, the total flux from root to leaves is 
\begin{equation}
\label{PhiN}
\Phi_N=\langle c\rangle \ln N ~.
\end{equation}
Fluctuations around the average are theoretically negligible in the $N\to\infty$ limit, but since they scale as  $\sqrt{\ln N}$, see Eq.~(\ref{v1}), 
fluctuations become negligible only for immensly large $N$.

\section{Discussion}
\label{disco}

Trees represent the least interconnected graphs while complete graphs are the most connected. Flows in complete graphs exhibit simple asymptotic behavior.  The maximal flux from an arbitrarily chosen source node to an arbitrarily chosen sink node scales as
\begin{equation}
\label{complete}
\langle s\rangle_N=\langle c\rangle N
\end{equation}
and fluctuations about this average are asymptotically negligible. To show this we first notice that for a complete graph with $N+1$ nodes, the flow from node 0 (source) to node $N$ (sink) cannot exceed $\sum_{1\leq j\leq N}Nc_{0j}$
which is asymptotically $\langle c\rangle N$ (fluctuations are of the order of $\sqrt{N}$). This is the upper bound and from a particular node $j$ it will be impossible to transmit flow $c_{0j}$ directly to the sink if $c_{0j}>c_{jN}$. However the node $j$ is connected to nodes $1,\ldots,j-1,j+1,\ldots N-1$ and in the large $N$ limit it will be possible to find the way to transfer the flow via those side routs.

For regular lattices, the problem seems analytically intractable. The simplest set-up is a two-dimensional square lattice with additional restriction that the flow is biased along the diagonal. More precisely, the root is $(0,0)$ and from every site $(i,j)$ with $i,j\geq 0$ the flow is to $(i+1,j)$ and $(i,j+1)$. The capacities are again chosen independently from the same distribution. A similar problem of directed polymers on the same lattice with random `energies' assigned to nodes has been solved by Krug and Halpin-Healy \cite{krug}. The difficulty in our case arises from the splitting of current at nodes. 

A related problem is the minimal cost flows on networks \cite{network}. In these models there is a cost associated with each edge and the cost of a flow is the sum of the costs of the edges it flows through. This problem was considered with random costs drawn from a common distribution \cite{cieplak, sam}. When the costs $\tau$ are taken from an exponential distribution $\exp(a\tau)$, the $a\to\infty$ limit is called the strong disorder limit. In this case the cost of a flow is dominated by its single most expensive edge, and the search for the cheapest flow becomes equivalent to our present study on trees: the search for the maximal flow.

\bigskip
\noindent{\bf Acknowledgments.} We thank Sameet Sreenivasan for comments. 
The Program for Evolutionary Dynamics at Harvard University is sponsored by Jeffrey Epstein. PLK acknowledges support from NSF grant CHE-0532969.

\appendix
\section{small $q$ series of $\rho(q)$}
\label{smallq}

Assume that we can write $f(q)$ and $\rho(q)$ as a series around $q=0$
\begin{equation}
f(q) = \sum_{i=0}^\infty \frac{f^{(i)}(0)}{i!} q^i ~,~~~ \rho(q) = \sum_{j=0}^\infty \frac{\rho^{(j)}(0)}{j!} q^j 
\end{equation}
where $\cdot^{(i)}$ denotes the $i$-th derivative of a function. The integrated probabilities then can be written as
\begin{equation*}
 F(q) = -\sum_{i=0}^\infty \frac{f^{(i-1)}(0)}{i!} q^{i} ~,~~~
 R(q) = -\sum_{j=0}^\infty \frac{\rho^{(j-1)}(0)}{j!} q^{j}
\end{equation*}
where formally we denote $f^{(-1)}(0)\equiv \rho^{(-1)}(0)\equiv -1$.
Then from \eqref{r-gen} we have
\begin{equation*}
H(q) = \sum_{n=0}^\infty q^n \sum_{i=0}^n \frac{f^{(i-1)}(0) \rho^{(n-i-1)}(0)}{i!(n-i)!}
\end{equation*}
from which we obtain
\begin{equation}
 -h^{(n)} (0) = (n+1)! \sum_{i=0}^{n+1} \frac{f^{(i-1)}(0) \rho^{(n-i)}(0)}{i!(n+1-i)!} ~.
 \label{ser-h}
\end{equation}
On the other hand, from \eqref{conv}
\begin{equation}
 \rho^{(k)}(0) = \sum_{j=1}^{k} h^{(j-1)}(0) h^{(k-j)}(0) ~.
 \label{ser-rho}
\end{equation}
Hence, with \eqref{ser-h} and \eqref{ser-rho}, we explicitly expressed $\rho^{(j)}(0)$ with lower derivatives of $\rho$, which recursion can be easily implemented on a computer. The limiting distribution is compared to simulation results in Fig.~\ref{totexpo} and \ref{totflat} (for the flat capacity distribution the result is valid only for $q<1$, as the radius of convergence is one for the flat distribution itself). It is also straightforward to obtain the leading terms of the asymptotic series by hand
\begin{equation*}
 \rho(q) = qf^2(0) + q^2 f(0)[f^2(0)+f'(0)] + \mathcal{O}(q^3) ~.
\end{equation*}

For binary hierarchical lattices the calculation is similar. Equation \eqref{ser-rho} is still valid, while one should replace \eqref{ser-h} by
\begin{equation*}
 -h^{(n)} (0) = (n+1)! \sum_{i, j, l \ge 0} \frac{f^{(i-1)}(0) f^{(j-1)}(0) \rho^{(l-1)}(0)}{i!j!l!} ~,
 \label{ser-h_hier}
\end{equation*}
where $i+j+l=n+1$ is required in the sum.

The above calculations can be straightforwardly extended to $b$-ary trees and $b$-ary 
hierarchical lattices.


\begin{thebibliography}{99}
\bibitem{note}
     Equations (\ref{d1})--(\ref{v1}) show that the relative magnitude of fluctuations 
     is $\mathcal{V_1}^{1/2}/\mathcal{D}_1\sim (\ln N)^{-1/2}$,  
     so in practice fluctuations can be ignored only for astronomically large recursive trees.
\end{thebibliography}
\end{document}